# Competition between Glassy Five-Fold Structures and Locally Dense Packing Structures Governs Two-Stage Compaction of Granular Hexapods

Rudan Luo,[1] Houfei Yuan,[3] Yi Xing,[3] Yeqiang Huang,[1] Jiahao Liu,[1] Wei Huang,[1] Haiyang Lu,[3] Zhuan Ge,[1] Yonglun Jiang,[1] Chengjie Xia,[4] Zhikun Zeng,[3,*] and Yujie Wang[1,2,3,†]

[1]*School of Physics, Chengdu University of Technology, Chengdu 610059, China*
[2]*State Key Laboratory of Geohazard Prevention and Geoenvironment Protection, Chengdu University of Technology, Chengdu 610059, China*
[3]*School of Physics and Astronomy, Shanghai Jiao Tong University, Shanghai 200240, China*
[4]*School of Physics and Electronic Science, East China Normal University, Shanghai 200241, China*

Using X-ray tomography, we experimentally investigate the structural evolution of packings composed of 3D-printed hexapod particles, each formed by three mutually orthogonal spherocylinders, during tap-induced compaction. We identify two distinct structural compaction mechanisms: an initial stage dominated by enhanced particle interlocking, which yields local mechanically stable structures through strong geometric entanglement, and a later stage characterized by the formation of dense polytetrahedral aggregates and a sharp increase in the number of five-ring motifs. The emergence of these five-fold symmetric structures indicates that, despite their highly concave geometry, hexapod packings can be effectively treated as hard-sphere-like systems and exhibit similar glass-like disordered configurations. The frustration between local mechanically stable structures and global glassy order suggests a universal organizational principle underlying the structure of uniform and isotropic disordered granular materials.

Granular matter, from sand dunes to construction aggregates, is ubiquitous in both natural processes and industrial applications [1]. In the absence of external driving forces, granular materials typically form stable, static packings due to their athermal nature and complex dissipative interactions among particles [2,3]. Packings of spherical granular particles are often regarded as prototypical disordered systems and share great structural similarities with thermal hard-sphere glassy systems [4,5]. A notable feature of these disordered systems is the emergence of dense tetrahedral configurations, where tetrahedra connect through shared faces to form polytetrahedral aggregates, representing a form of medium-range structural order [6]. These polytetrahedral aggregates can be characterized as consisting of N-ring structures, which are specific tetrahedral clusters made up of $N$ tetrahedra sharing a common edge [7]. Recent works have demonstrated that these polytetrahedral structures are responsible for packing arrangements [8,9], plastic events [10,11], and the overall mechanical properties of the systems [12]. In contrast, packings composed of highly non-convex particles, such as U-shaped staples and hexapod-shaped particles, exhibit markedly different structural properties [13-16]. Their concave geometry enables the formation of local interlocked structures through mutual hindrance between recesses and protrusions, effectively restricting particle motion [17]. This interlocking motif, by nature, deviates from conventional glass-like disordered configurations and induces "geometric cohesion" within the system, which allows granular piles formed by non-convex particles to exhibit high mechanical stability with angles of repose approaching 90° [18,19], a property typically associated with cohesive materials. Despite the prevailing belief that the interlocking effect arises from mutual "hooking" between non-convex particles, no systematic experimental study has yet explored their microscopic packing configurations. Furthermore, while packings of spherical particles under external perturbations such as vibration or shear exhibit slow, glass-like compaction dynamics [20-22], it remains unclear whether concave particle systems display similar relaxation behavior. Understanding how local structural organizations, including the interlocking structures, affect the compaction dynamics and mechanical properties of concave particle packings is critical for practical applications, including powder processing, additive manufacturing, and architected granular materials [18].

In this Letter, we use hexapod-shaped particles composed of three mutually orthogonal spherocylinders to investigate the compaction process of concave particle systems. We generate granular hexapod packings across a wide range of volume fractions through tapping and analyze their microscopic structures during compaction and at steady states using X-ray tomography. Our results reveal the coexistence of two distinct structural motifs in the systems: local interlocked clusters, typically comprising two or three geometrically entangled particles, and globally disordered configurations formed by polytetrahedral aggregates resembling those in spherical particle packings. These structures form at different length scales and compete throughout the compaction process, giving rise to a two-stage compaction behavior.

The concave particles used in this study are 3D-printed (ProJet MJP 2500 Plus, resolution of 0.032 mm) plastic six-fold symmetric hexapods, each composed of three mutually orthogonal spherocylinders with hemispherical caps, as shown in Fig. 1(a). The diameter of the spherocylinders is $d = 4$ mm, and the length of the cylindrical part is $h = 16$ mm, resulting in an aspect ratio $\alpha = (h+d)/d = 5$. To enable individual particle identification during imaging, a spherical hole with a diameter of 2 mm is designed at the center of the grain, with six additional 2.6 mm spherical holes located at the center of each ball cap [Fig. 1(a)]. We generate disordered granular packings containing more than 1,400 hexapods in a cylindrical container with an inner diameter of 150 mm and a packing height of roughly 160 mm. To prepare a reproducible initial loose packing, we insert a thinner cylindrical tube into the container and pour hexapods through it. We then slowly withdraw the tube vertically, allowing the hexapods to settle gently in the container under gravity. Subsequently, packings with a wide range of densities can be realized by tapping the system using a mechanical shaker with tap intensities $\Gamma = 4.6 \sim 45g$, where $g$ is the gravitational acceleration constant. Each tap consists of a 30 ms half-sine pulse followed by a 1.5 s interval to allow the system to settle. The system is tapped for 100~100,000 times, depending on $\Gamma$, to reach their corresponding steady states.

The evolution of the packing structures under tapping is obtained using a medical CT scanner (UEG Medical Group Ltd., 0.2 mm spatial resolution). Following similar image processing procedures as previous studies [7,23], the centroid and orientation of each hexapod are determined with uncertainties of less than $0.01d$ and 0.1 degrees, respectively. Specifically, the particle center is identified through the central hole of the hexapod, while orientation is determined from the three vectors connecting the paired holes at the ends of each spherocylinder. In subsequent analyses, we only include hexapods located at least $3d$ away from the container walls and the free top surface, resulting in approximately 700 particles in the bulk region. A representative packing structure is shown in Fig. 1(b).

To investigate the packing structures during compaction, we first analyze the evolution of packing fraction $\phi = v_p / \langle v_{voro} \rangle$ as a function of tapping number $t$ under varying $\Gamma$, where $v_p$ and $v_{voro}$ represent the respective volumes of the hexapod particles and their associated Voronoi cell obtained via the set Voronoi constructions [24], and $\langle ... \rangle$ represents the average of all hexapods in the bulk region. As shown in Fig. 1(c), although hexapod packings display significantly lower $\phi$ due to geometric interlocking, they gradually compact and reach corresponding steady states in a manner similar to spherical particles [25]. We quantify this process by fitting the compaction curves with the Kohlrausch-Williams-Watts (KWW) form:

$$\phi(t) = \phi_f - (\phi_f - \phi_0)\exp\left[-\left(\frac{t}{\tau}\right)^\beta\right], \qquad (1)$$

where $\tau$ is the relaxation time, $\beta$ is the stretching exponent, and $\phi_0$ and $\phi_f$ denote the initial and steady-state volume fractions, respectively. Figure 1(d) shows the $\Gamma$-dependencies of $\phi_f$ and $\tau$. Both $\phi_f$ and $\tau$ increase with decreasing $\Gamma$ as expected, but a distinct kink is observed at $\Gamma = 18g$ (or $\phi_c = 0.39$). This anomaly is further supported by the behavior of the fitted stretching exponent $\beta$, which remains constant for $\Gamma > 18g$ but decreases rapidly from 0.8 to 0.5 for $\Gamma < 18g$ [inset of Fig. 1(d)], implying a more heterogeneous relaxation process involving multiple mechanisms. Together, these observations suggest the presence of a transition between two distinct phases.

The compaction mechanisms of hexapod systems can be inferred by comparing the steady-state packing structures under different $\Gamma$. Figure 2(a) shows the radial pair correlation function $g(r)$ for steady-state packings under varying $\Gamma$. We observe that $g(r)$ displays a small hump around $r = 2.1d$ and a primary peak at $r = 3.1d$, indicating two-layer neighboring structures. Beyond the particle end-to-end distance $h + d = 5d$, $g(r)$ rapidly decays to 1, indicating the absence of long-range correlation in the packings. This behavior contrasts with that of spherical particle packings, where $g(r)$ exhibits fluctuations at larger distances [26]. The evolution of $g(r)$ peaks show similar kinks that are consistent with the anomalies observed in $\phi$ and $\tau$. For $\Gamma > 18g$ or $\phi_c < 0.39$, the hump around $r = 2.1d$ becomes more pronounced as $\phi$ increases, and the peak position around $r = 3.1d$ shifts slightly to smaller values

[Fig.2(b)]. These trends indicate that during the initial stage of compaction, the number of first-layer neighbors at $r = 2.1d$ increases, accompanied by a slight reduction in the distance to second-layer neighbors. However, both trends slow down abruptly for $\phi > 0.39$, suggesting that an additional compaction mechanism not captured by $g(r)$ is at work.

To better understand the two-layer neighboring structures, we further analyze the orientational properties of the hexapod particles. Overall, hexapods are uniformly distributed across different orientations (see Supplemental Material [27]). To quantify the orientational correlation induced by particle entanglement, we define the angular correlation function $\Omega = \max\left(\sum_{i=1}^{3}|\vec{n}_{pi} \cdot \vec{n}_{qi}|/3\right)$, where $\vec{n}_{pi}$ and $\vec{n}_{qi}$ denote the unit vectors of the three spherocylinders ($i$ = 1, 2, 3) of hexapods $p$ and $q$. For maximally interlocked structures, the six arms of adjacent particles are parallel with $\Omega = 1$ [inset of Fig. 2(c)], while randomly oriented pairs yield an average value of $\Omega \approx 0.825$. The angular correlation function $\left\langle \Omega_{|\mathbf{r}_p - \mathbf{r}_q|=r} \right\rangle$ is shown in Fig. 2(c). The hump in $g(r)$ near $2.1d$ corresponds to significantly interlocked structures with $\Omega \sim 0.9$ [left inset of Fig. 2(a)], while the primary peak of $g(r)$ at $3.1d$ represents local configurations where an arm of a neighboring particle inserts deep into the center particle with only a slight increase in orientational correlation [right inset of Fig. 2(a)]. We also calculate the average value of $\Omega$ between a center particle and its nearest neighbor during compaction [Fig. 2(d)]. $\Omega$ increases for $\phi < 0.39$ and plateaus for denser packings. Combined with the information obtained from $g(r)$, this indicates that, at least for the first stage, compaction happens by enhanced particle interlocking.

Next, we characterize the interlocked structures within the packings. Specifically, hexapods with interparticle distances $d_{inter} < 2.5d$ and $\Omega > 0.85$ are considered interlocked pairs, which further connect to form interlocked particle clusters (IPCs). Figure 3(a) shows the IPCs and their corresponding networks in loose ($\phi = 0.372$) and dense ($\phi = 0.427$) packings. Within IPCs, particles are tightly bound through geometric entanglement, resembling orientation-specific short-range attractions in patchy particle systems [28]. However, due to the steric constraints imposed by the hexapod shape, strongly interlocked neighbors can coexist only in opposite directions around the central particle, which inhibits the further spatial growth of IPCs. Consequently, most IPCs remain small, typically involving only two or three particles. As shown in Fig. 3(b), the fraction of interlocked particles $P_{\text{IPC}}$ increases for $\phi < 0.39$ and saturates for denser packings, consistent with the evolution of $g(r)_{r=2.1d}$. However, the average center-to-center distance between the nearest IPCs $d_{\text{cluster}}$ continues to decrease for $\phi > 0.39$ [see Fig. 3(c)], suggesting that an additional mechanism becomes dominant in further densifying the system.

In order to elucidate the second-stage compaction mechanisms beyond local interlocking, we analyze the contact properties of the systems [29]. Unlike spherical particles, hexapods can have multiple contact points between pairs. We therefore distinguish two parameters: the average coordination number $Z$, defined as the average number of contacting neighbors per particle, and the average number of contacting points per particle $N_c$ [30]. The dependencies of $Z$ and $N_c$ on $\phi$ are shown in Fig. 4(a) and its inset. Both $Z$ and $N_c$ have a weak kink at $\phi = 0.39$, after which they increase more rapidly, suggesting that denser compaction is driven by the formation of additional contact neighbors. To probe this behavior in more detail, we classify the contacting particles into two categories based on their center-to-center distance $d_{\text{inter}}$. Contacting particles with $d_{\text{inter}} < L/2 = 2.5d$ are classified as first-layer neighbors, corresponding to strongly interlocked neighbors around $g(r)_{r=2.1d}$, while those with $d_{inter} \geq L/2$ are designated as second-layer neighbors, associated with the primary peak of $g(r)_{r=3.1d}$. We note that the results are insensitive to the specific threshold choices in determining the two layers. As shown in the inset of Fig. 4(b), the number of first-layer contacting neighbors $Z_1$ increases with $\phi$ but the growth slows down for $\phi > 0.39$, consistent with the earlier conclusion that particle interlocking dominates compaction for $\phi < 0.39$ (see Supplemental Material [27] for more details). In contrast, the number of second-layer contacting neighbors $Z_2$ rises rapidly once $\phi$ exceeds 0.39, indicating that, beyond this point, compaction proceeds primarily through the rearrangement of second-layer neighbors to increase Z.

The increase of Z during the second stage of compaction mirrors similar phenomena observed in spherical particle packings [31]. To further characterize the hexapod packing structures, we employ polytetrahedral analysis, a common approach for examining the structural features of disordered spherical packings. Specifically, Delaunay tessellation partitions the packing structure into non-overlapping tetrahedra, with each tetrahedron defined by four neighboring particles whose centers form the vertices. Following previous studies [5,6], we calculate the tetrahedral order parameter $\delta = \frac{e_{max}}{r_{peak}} - 1$, where $e_{max}$ is the length of the longest tetrahedron edge and $r_{peak}$ denotes the particle distance at the first peak of $g(r)$, which is the mean particle diameter in spherical packings and corresponds to $3.1d$ in our hexapod packings. We note that $\delta$ is insensitive to short edges formed by interlocked neighbors and thus selectively probes non-interlocked structural order. Figure 5(a) shows the average values $\langle \delta \rangle$ for steady-state systems at varying $\phi$. We observe a monotonic decrease in $\langle \delta \rangle$ with increasing $\phi$, consistent with trends in spherical particle packings [31]. Quasi-regular tetrahedra are defined as tetrahedra whose shapes closely approximate regular tetrahedra, with $\delta$ smaller than a specified threshold. In previous studies of spherical packings, this threshold is approximately 0.25, corresponding to the distance where $g(r)$ first decays to 1. For our hexapod packings, quasi-regular tetrahedra are defined as those with $\delta < \frac{4.2d}{3.1d} - 1 \approx 0.35$, as $g(r)$ first decays to 1 at 4.2$d$. As shown in Fig. 5(b), the fraction of quasi-regular tetrahedra $P(\delta < 0.35)$ increases significantly with $\phi$. Both $\langle \delta \rangle$ and $P(\delta < 0.35)$ vary linearly with $\phi$, without any abrupt kinks around $\phi = 0.39$, indicating that the second-stage compaction is not driven by the structural evolution of individual tetrahedra.

Instead, we investigate the aggregation behavior of these tetrahedra by analyzing networks formed by quasi-regular tetrahedra. We construct networks by connecting face-adjacent quasi-regular tetrahedra and visualize their centers and linkages in Fig. 5(c). In denser systems, quasi-regular tetrahedra increase in number and aggregate into *N*-ring structures, groups of tetrahedra sharing a common edge and being coplanar between neighboring members [6,7]. These *N*-ring structures were originally developed to describe the potential ideal glass state of hard-sphere systems, where the five-ring is often considered the disclination-free ground state structure and a hallmark of medium-range order [7,32]. The emergence of five-fold symmetric structures in hexapod packings suggests that, despite their highly concave shape, such centrally symmetric particles can be effectively treated as spherical ones with $r_{peak}$ serving as an effective particle diameter, giving rise to similar glass-like structures [Inset of Fig. 5(d)]. Interestingly, the number of five-ring structures increases sharply only when $\phi > 0.39$ [Fig. 5(d)], which coincides exactly with the kink observed in the compaction curve. This sudden prevalence of five-ring structures signals a transition from an initial loose, entangled state to a more densely packed, glass-like state, resembling that from an open-network, gel-like phase to a compact, hard-sphere-like glassy phase observed in gels or attractive glasses [33,34]. This interpretation is supported by the observation that $\langle \delta \rangle$ for our hexapod systems after the kink correspond to those of spherical packings with $\phi$ ranging from 0.57 to 0.62, roughly matching the range from the random loose packing (RLP) state to the random close packing (RCP) state of spherical granular systems [31,35]. This correspondence is further reinforced by the observation that the compaction dynamics of hexapod systems follow the Adam–Gibbs (AG) relation similar to spherical granular packings in this volume fraction range [27].

In summary, we have investigated the structural characteristics of hexapod packings under different tap intensities, providing insights into the underlying mechanisms governing their compaction process. We identify two distinct compaction pathways: the first driven by enhanced particle interlocking, forming local mechanically stable structures with strong geometric entanglement, and the second characterized by the emergence of dense polytetrahedral arrangements, indicative of globally disordered, glass-like configurations. These two structural motifs that form at different length scales compete throughout the compaction process, reminiscent of the concept of "geometrical frustration" [8]. We propose that this competition between local mechanically stable structures and globally glassy configurations with five-fold symmetry is a generic feature of disordered packings irrespective of their particle shape. This phenomenon closely resembles Miracle's concept, which describes glassy packing structures as comprising different structural orders at various length scales, where the order is medium-range crystalline ordering in his case and quasi-crystalline or five-fold symmetric structures in our case [36]. These insights can greatly enhance our knowledge of structural organization in uniform and isotropic disordered granular packing systems.

We would like to acknowledge discussions with W. Kob. The work is supported by the National Natural Science Foundation of China (No. 12274292) and the Space Application System of China Manned Space Program (KJZ-YY-NLT0504). Z.Z. acknowledges support from the National Natural Science Foundation of China (No. 123B2060).

Corresponding author.

*zzk97115_kenny@sjtu.edu.cn

†yujiewang@sjtu.edu.cn

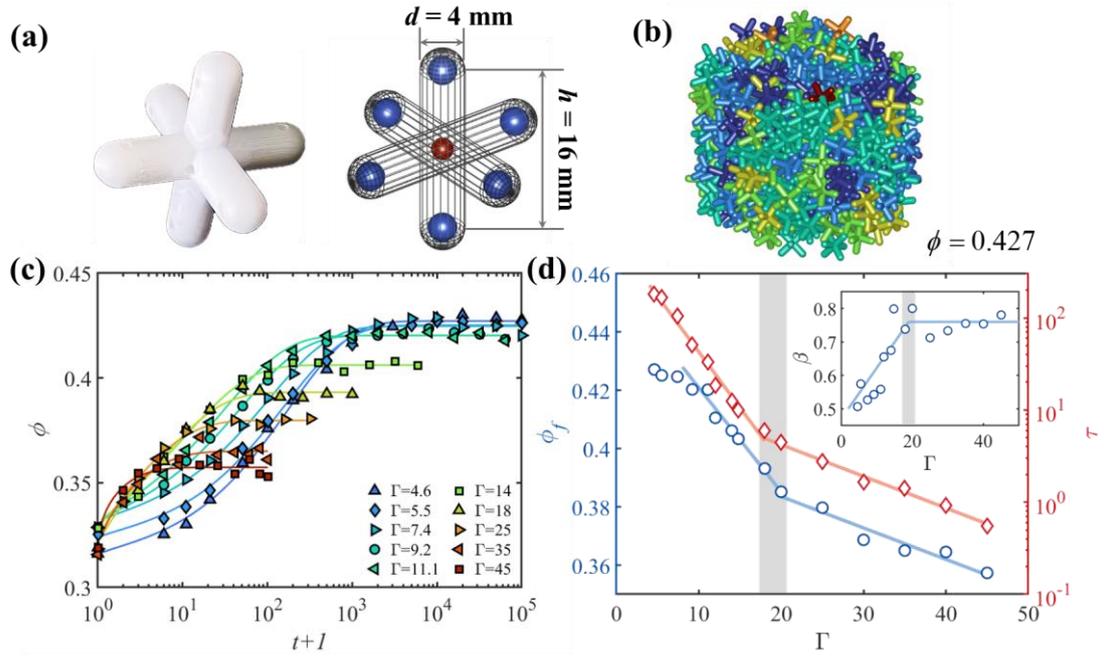

FIG. 1. (a) Photograph of a hexapod-shaped particle (left) and a schematic diagram of its internal hollow holes that facilitate particle identification in CT images (right). (b) Structures of a hexapod packing with $\phi = 0.427$. Particles are colored from blue to red based on the distance to their nearest neighbor, from small to large. (c) Volume fraction $\phi$ as a function of tap numbers $t$ for different $\Gamma$. The solid curves represent the KWW fitting. (d) Steady-state volume fractions $\phi_f$ and relaxation time $\tau$ as a function of $\Gamma$. Inset: the fitted stretched exponent $\beta$ as a function of $\Gamma$. The solid lines are guide to the eye.

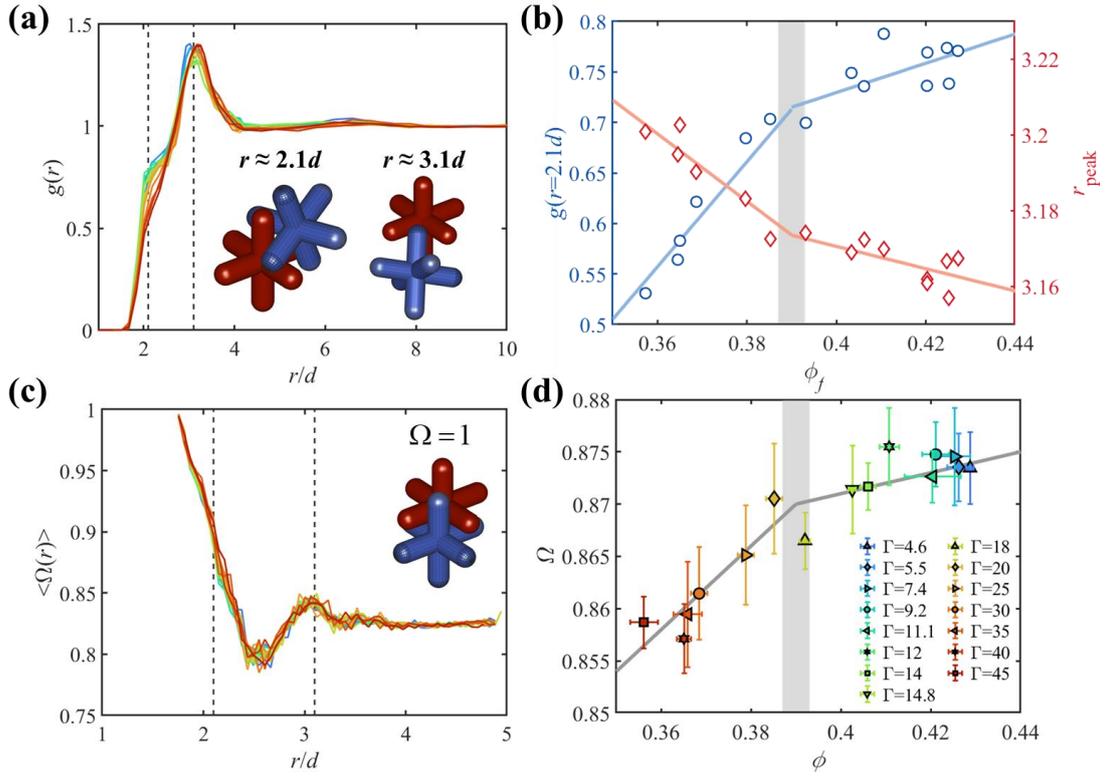

FIG. 2. (a) Pair correlation function $g(r)$ for different $\Gamma$. Inset: schematic diagram of particle configurations for $r = 2.1d$ (left) and $3.1d$ (right). (b) $g(r = 2.1d)$ and the peak location $r_{peak}$ as a function of $\phi_f$. (c) Angular correlation function $\langle \Omega(r) \rangle$ for different $\Gamma$. Inset: schematic diagram of the maximum interlocked configuration with $\Omega = 1$. (d) Averaged $\Omega$ between central particles and their nearest neighbors versus packing fraction $\phi$. The solid lines are guide to the eye.

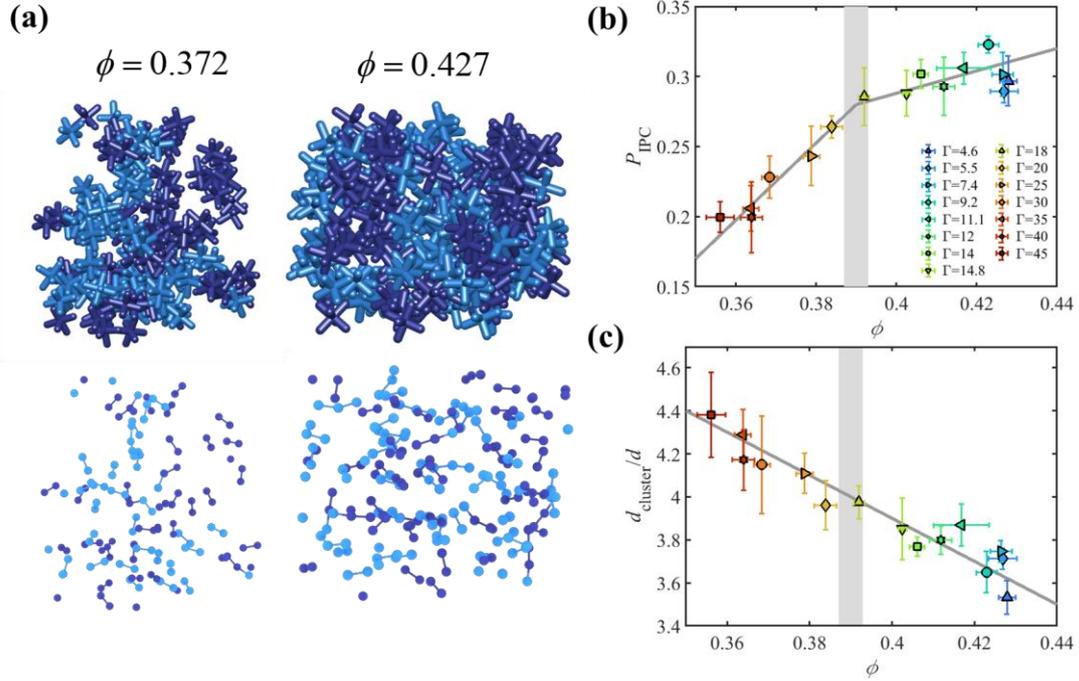

FIG. 3. (a) Interlocked particle clusters (upper) and their corresponding networks (lower) in hexapod packings with $\phi = 0.372$ and 0.427, respectively. Hexapods with interparticle distances less than $2.1d$ are colored by dark blue. (b) Fraction of interlocked particles $P_{IPC}$ as a function of $\phi$. (c) Average distance between the nearest IPCs $d_{cluster}$ as a function of $\phi$. The solid lines are guide to the eye.

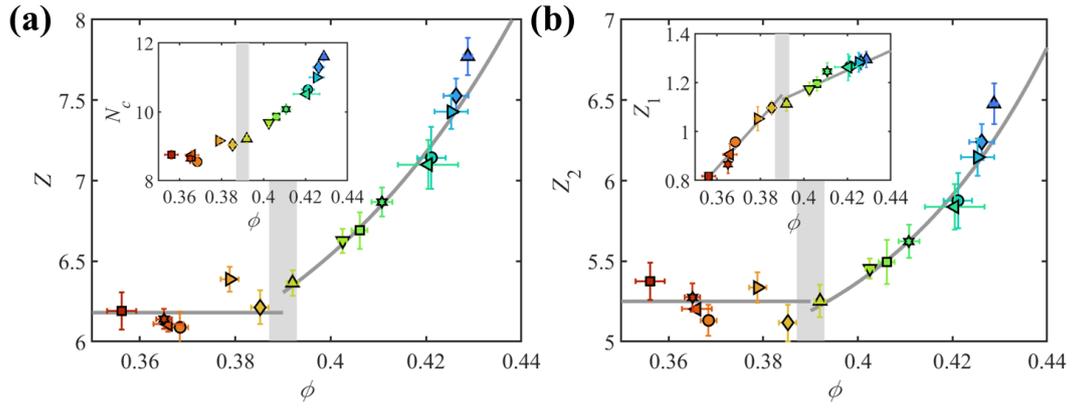

FIG. 4. (a) Coordination number $Z$ and contact point number $N_c$ (inset) as a function of $\phi$. (b) Number of first- and second-layer contacting neighbors, $Z_1$ (inset) and $Z_2$ respectively, as a function of $\phi$. The solid curves are guide to the eye.

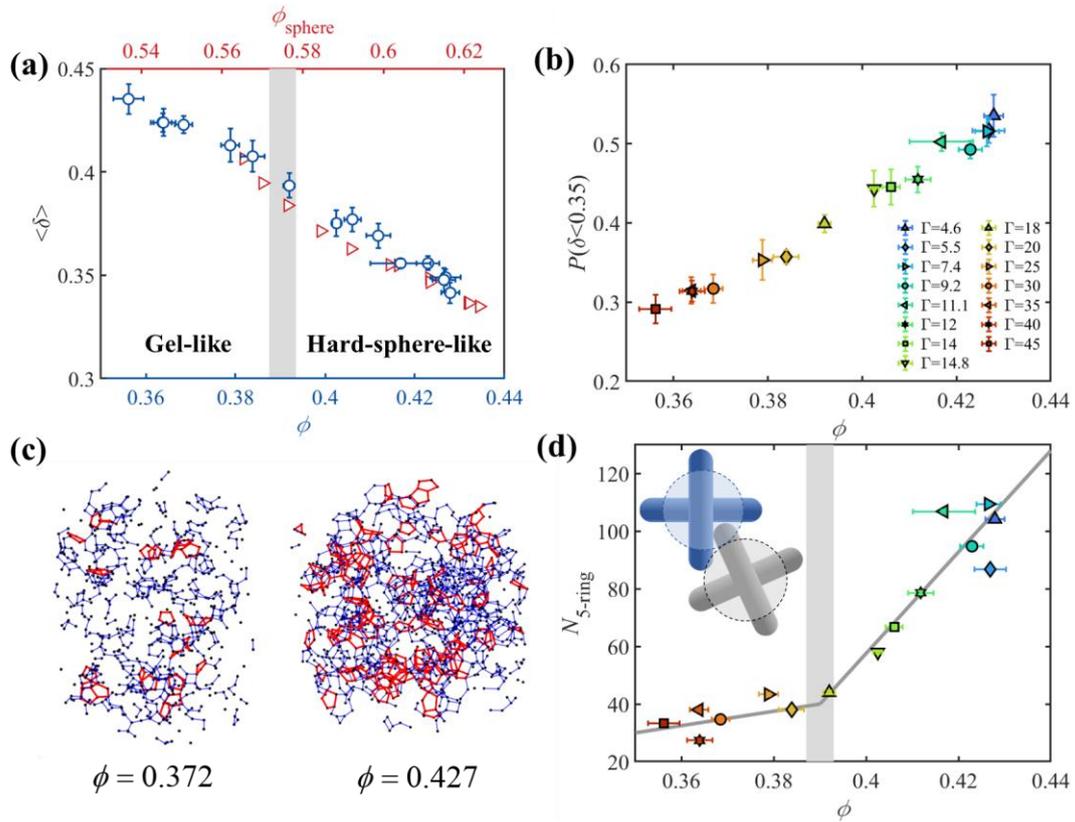

FIG. 5. (a) The averaged value of $\delta$ for steady-state hexapod (blue) and spherical particle (red) packings across different packing fractions $\phi$ and $\phi_{sphere}$. The data for spherical particle packings is from Ref. [31]. (b) The fraction of quasi-regular tetrahedra $P(\delta<0.35)$ as a function of $\phi$. (c) Networks of polytetrahedral aggregates in hexapod packings with $\phi=0.372$ and 0.427, respectively. Black dots represent the quasi-regular tetrahedra and blue lines connect two tetrahedra sharing a common face. The five-ring structures are marked in red. (d) The number of five-ring structures as a function of $\phi$. Inset: schematic illustration showing that concave particles can effectively be treated as spherical ones. The solid lines are guide to the eye.

# Supplemental Material for

# Structural Mechanisms of Two-Stage Compaction in Granular Hexapod Packings under Tapping

Rudan Luo,[1] Houfei Yuan,[3] Yi Xing,[3] Yeqiang Huang,[1] Jiahao Liu,[1] Wei Huang,[1] Haiyang Lu,[3] Zhuan Ge,[1] Yonglun Jiang,[1] Chengjie Xia,[4] Zhikun Zeng,[3,*] and Yujie Wang[1,2,3,†]

[1] School of Physics, Chengdu University of Technology, Chengdu 610059, China
[2] State Key Laboratory of Geohazard Prevention and Geoenvironment Protection, Chengdu University of Technology, Chengdu 610059, China
[3] School of Physics and Astronomy, Shanghai Jiao Tong University, Shanghai 200240, China
[4] School of Physics and Electronic Science, East China Normal University, Shanghai 200241, China

## 1. Overall Orientation Distributions.

To assess whether the system exhibits orientational isotropy, we calculate the probability distribution functions (PDFs) for the angles $\theta_z$ between the three orthogonal axes of each hexapod particle and the vertical (gravitational) direction, as shown in Fig. S1. The results reveal a uniform distribution across all angles, indicating that particle orientations are homogeneously distributed with respect to gravity, and that no preferred orientation emerges throughout the system at steady state.

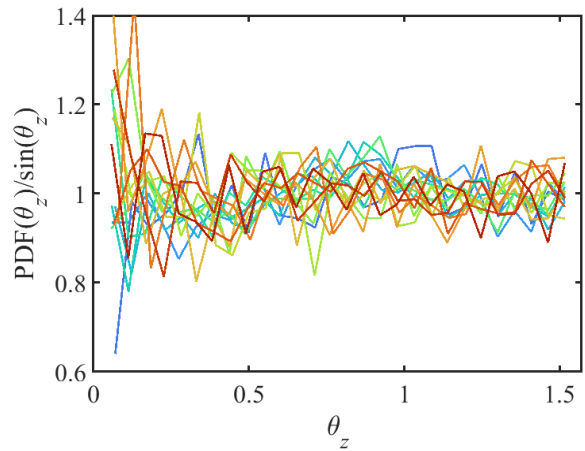

FIG. S1. Normalized PDFs of the angles between three hexapod particle orientations and the vertical direction $\theta_z$ for steady-state hexapod packings at different Γ.

**2. Contact Detection.**

Contacts between hexapods are identified based on the surface-to-surface distance between particles. Using the positions and orientations of the hexapod particles, we first extract the central axes of their three constituent spherocylinders, defined as line segments connecting the centers of the hemispherical caps at each end [Fig. S2(a)]. Ideally, two spherocylinders are considered in contact when the minimum distance $r_{arm}$ between their axes equals the spherocylinder diameter $d = 4$ mm. However, due to limitations in CT resolution, imaging artifacts, and segmentation errors, the distribution of surface-to-surface distances $\Delta r = r_{arm} - d$ between contacting particles exhibits a Gaussian-like profile [see Fig. S2(b)]. Moreover, a shoulder appears at $\Delta r > 0$, arising from particles that are spatially close but not in actual contact.

To determine the average coordination number $Z_c$, we adopt a threshold value $\delta_{th}$ for the surface-to-surface distance $\Delta r$, such that particles with $\Delta r < \delta_{th}$ are considered to be in contact. The dependence of $Z_c$ with different $\delta_{th}$ is shown in Fig. S2(c), and can be well described by a superposition of an error function capturing the Gaussian-like distribution of $\Delta r$ for contacting particles (red curve), and a linear function denoting contributions from non-contacting neighbors (purple line). By selecting a threshold $\delta_c$ that accurately isolates the error-function component associated with true contacts, we can determine both the average coordination number $Z_c$ of the system and the corresponding surface-to-surface distance threshold $\delta_c$ for identifying particle contacts.

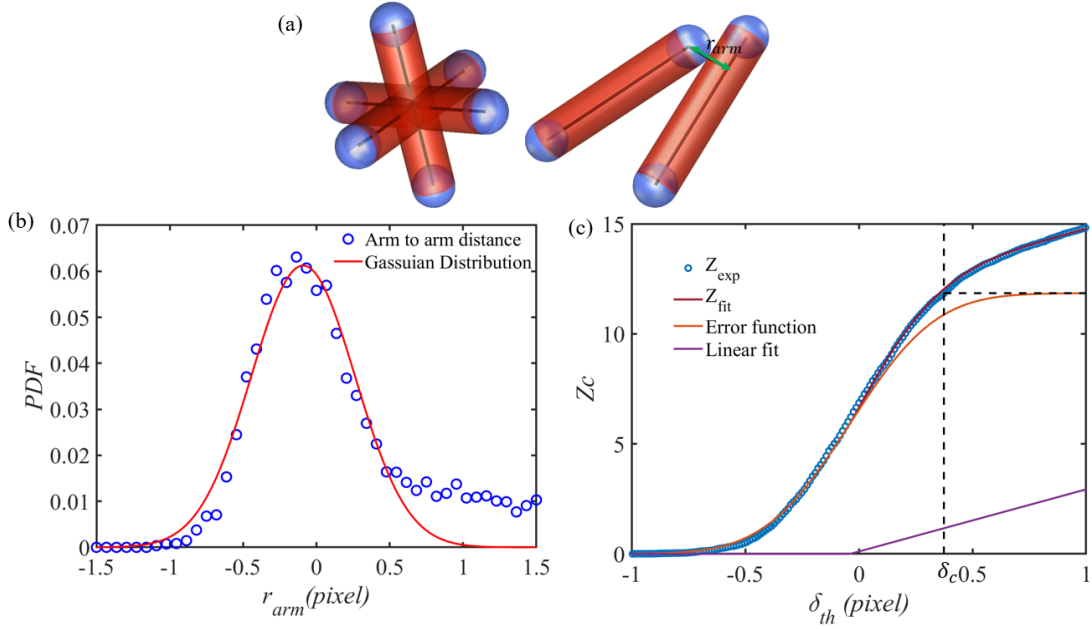

Fig. S2. (a) Schematic diagram of the three spherocylinder axes of a hexapod particle and the corresponding arm-to-arm distance $r_{arm}$. (b) Probability distribution function of surface-to-surface distance $\Delta r$ among neighboring particles. (c) Relationship between the average coordination number $Z_c$ and the contact threshold $\delta_{th}$, fitted by a superposition of an error function (red curve) and a linear function (purple line).

## 3. Degree of Interlocking and Contact Distance.

The degree of interlocking can be quantitatively defined as $\lambda = 1 - \langle d_{inter} \rangle / L$, where $\langle d_{inter} \rangle$ is the average distance between the centers of two contacting hexapods, and $L$ denotes the hexapod's end-to-end distance. A higher value of $\lambda$ indicates stronger geometric entanglement, as interparticle distances decrease. The contact distance $d_c$ is defined as the average distance from the contact point along the contacted spherocylinder to the center of the hexapod. During compaction, $\lambda$ gradually increases and $d_c$ continuously decreases for

$\phi < 0.39$, suggesting enhanced interlocking among neighboring particles [Figs. S3(a) and 4(b)]. However, for $\phi > 0.39$, both $\lambda$ and $d_c$ evolve significantly slower as the system continues to compact, consistent with the saturation observed in the fraction of interlocked particles. We note that there exists a slight decrease in $\lambda$ for $\phi > 0.42$, showing that the rearrangement of second-layer configurations competes with the further enhancement of particle interlocking.

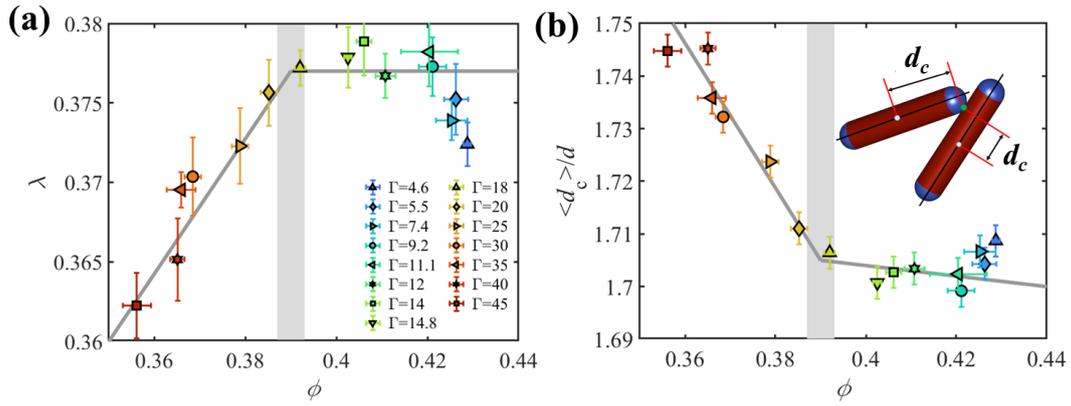

FIG. S3. (a) Degree of interlocking $\lambda$ and (b) contact distance $d_c$ as functions of $\phi$ for different $\Gamma$. Inset: schematic diagram of two spherocylinders in two contacting hexapods, showing how $d_c$ is measured. The solid lines are a guide to the eye.

## 4. Edwards Ensemble and Adam-Gibbs Relation.

Since hexapod packings exhibit disordered configurations similar to those of spherical particles, it is natural to explore their thermodynamic behavior using the Edwards ensemble framework. In analogy with granular spheres, we calculate the Voronoi cell volume variance var($v$) at different volume fractions [inset of Fig. S4(a)], where $v_p$ is set as unity for simplicity. The data is then fitted using a quadratic polynomial $\text{var}(v) = 4.7572 - 21.3573\phi + 24.2158\phi^2$ [black curve in the inset of Fig. S4(a)] to obtain an analytical expression. The compactivity $\chi$, which acts as an effective temperature for granular packings, can be determined by the

fluctuation method:

$$\frac{1}{\chi(\phi)} - \frac{1}{\chi^r} = \int_{\phi^r}^{\phi} \frac{d\varphi}{\varphi^2 var(v)}, \quad (S1)$$

where $\phi^r$ is the packing fraction of the reference state with infinite compactivity, which is set as $\phi^r = 0.39$ in our study, since disordered configurations only dominate beyond this threshold. Figure S4(a) shows $\chi^{-1}$ calculated via Eq. (S1). The configurational entropy $S(\phi)$ of the system can be obtained using another thermodynamic equation:

$$S(\phi) - S_{RCP} = \int_{\phi}^{\phi_{RCP}} \frac{d\varphi}{\varphi^2 \chi(\varphi)}. \quad (S2)$$

Here, we empirically assume $S_{RCP} \approx 1.1$ as the Shannon entropy. Using the compactivity and entropy values obtained within the Edwards volume ensemble, we test the Adam-Gibbs (AG) relation in our hexapod systems, which links structural relaxation time $\tau$ to the product $\chi S$:

$$\ln \frac{\tau}{\tau_0} \propto \left(\frac{1}{\chi S}\right). \quad (S3)$$

The dependence of $\tau$ on $\chi$ and $S$, along with the fit to Eq. (S3), is shown in Fig. S4(b), showing good agreement. This result demonstrates that the relaxation dynamics of hexapod particles in the second compaction stage resemble those of spherical particle systems and can, in principle, be described using thermodynamic models developed for glasses. Additionally, the AG relation exhibits two distinct branches across the turning point, indicating a possible change in entropy production mechanisms between different regimes.

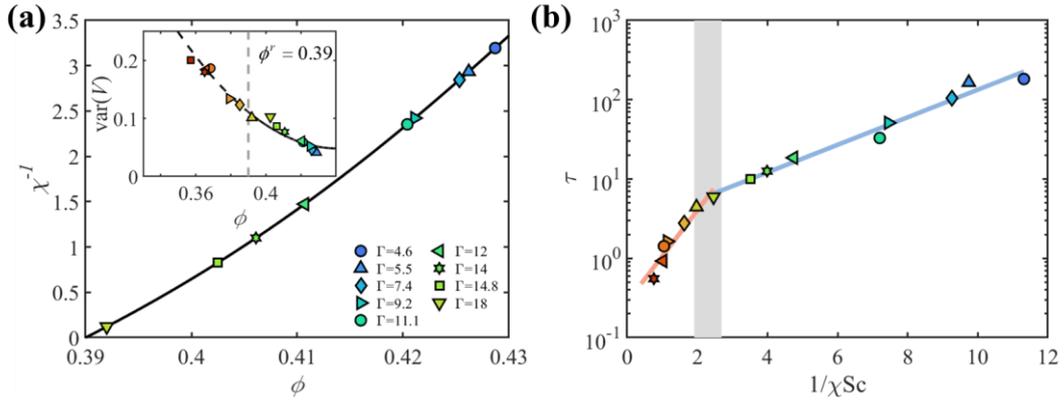

FIG. S4. (a) Volume fluctuation $\text{var}(V)$ as a function of $\phi$. The solid curve is a quadratic polynomial fit. Inset: inverse of compactivity $\chi^{-1}$ as a function of $\phi$ calculated via the fluctuation relation method. (c) Relaxation time $\tau$ versus $1/\chi S$ and fitting according to Eq. (S3).